\documentclass[journal,final]{IEEEtran}
\usepackage{color}
\usepackage{multirow}
\usepackage{graphicx}
\usepackage{epstopdf}
\usepackage[cmex10]{amsmath}
\usepackage{amssymb}

\usepackage{multirow}
\usepackage{amsthm}
\usepackage{cite}
\usepackage{flushend}
\usepackage{subcaption}
\usepackage{balance}

\begin{document}

\title{Metasurface-Coated Devices: A New Paradigm for Energy-Efficient and 
Secure 6G Communications}
\author{ Theodoros~A.~Tsiftsis, Constantinos Valagiannopoulos, Hongwu Liu,\\ Alexandros-Apostolos A. Boulogeorgos, and Nikolaos I. Miridakis
\thanks{T. A. Tsiftsis is with the School of Intelligent Systems Science \& Engineering, Jinan University, Zhuhai Campus, Zhuhai 519070, China (e-mail: theo\_tsiftsis@jnu.edu.cn).}
\thanks{C. Valagiannopoulos is with the Department of Physics, Nazarbayev University, Nur-Sultan KZ-010000, Kazakhstan
(e-mail: konstantinos.valagiannopoulos@nu.edu.kz).}
\thanks{H. Liu is with the School of Information Science and Electrical Engineering, Shandong Jiaotong University, Jinan 250357, China (e-mail: liuhongwu@sdjtu.edu.cn).}
\thanks{A.-A. A. Bouleogeorgos is with the Department of Digital Systems, University of Piraeus, Piraeus 18534, Greece. (e-mail: al.boulogeorgos@ieee.org).}
\thanks{N. I. Miridakis is with the Institute of Physical Internet and School of Intelligent Systems Science \& Engineering, Jinan University, Zhuhai Campus, Zhuhai 519070, China, and also with the Dept. of Informatics and Computer Engineering, University of West Attica, Aegaleo 12243, Greece (e-mail: nikozm@uniwa.gr).}
}



	\maketitle

	\begin{abstract}
	The sixth-generation (6G) era comes with the challenge
	of offering highly energy-efficient and autonomous communications
	securely. In this direction, we report energy efficiency
	(EE), energy harvesting (EH), and secure performance by
	employing \textit{power-collecting metasurface-coated devices} capable
	of supporting ultra-low-power (ULP) transmissions. Contrary
	to reconfigurable intelligent surfaces (RIS), where the reflected
	signal can be combined at the receiver by being treated as transmitted
	from a relay, the proposed metasurface claddings can be
	deployed at either or both the transmitter and receiver. The passive
	metasurface-coated devices can achieve ultra-high EE and EH
	besides the signal detection, combined with an enhanced secrecy
	rate at the legitimate user and/or improved spying capabilities
	of the eavesdroppers under ULP transmission. To quantify their
	efficiency, we provide a holistic model for the utilization of the
	metasurface shells. Building upon the aforementioned model,
	preliminary results are presented that reveal the
	unprecedented superiority of the proposed concept compared to the
	RIS paradigm. Additionally, we enumerate the main advantages of the new
	concept and define its role in the 6G era. Finally, possible
	research directions are discussed. 
\end{abstract}


\IEEEpeerreviewmaketitle

\section{Introduction}
The fifth-generation (5G)  of mobile communication networks has already been rolled out in many countries and currently provides promising services to billions of smart devices aiming at improving the civilians' lives. Nevertheless, the unprecedentedly increasing usage of smart devices driven by the next-generation internet-of-things (NG-IoT) applications will inevitably lead to an explosive growth of wireless data traffic. Apparently, the aforementioned growing demand of data-hungry 5G service has already overloaded existing cellular networks and appears as a bad sign for new deployments \cite{Traffic}.
\begin{figure*}
	\centering
	\includegraphics[keepaspectratio,width=6.5in]{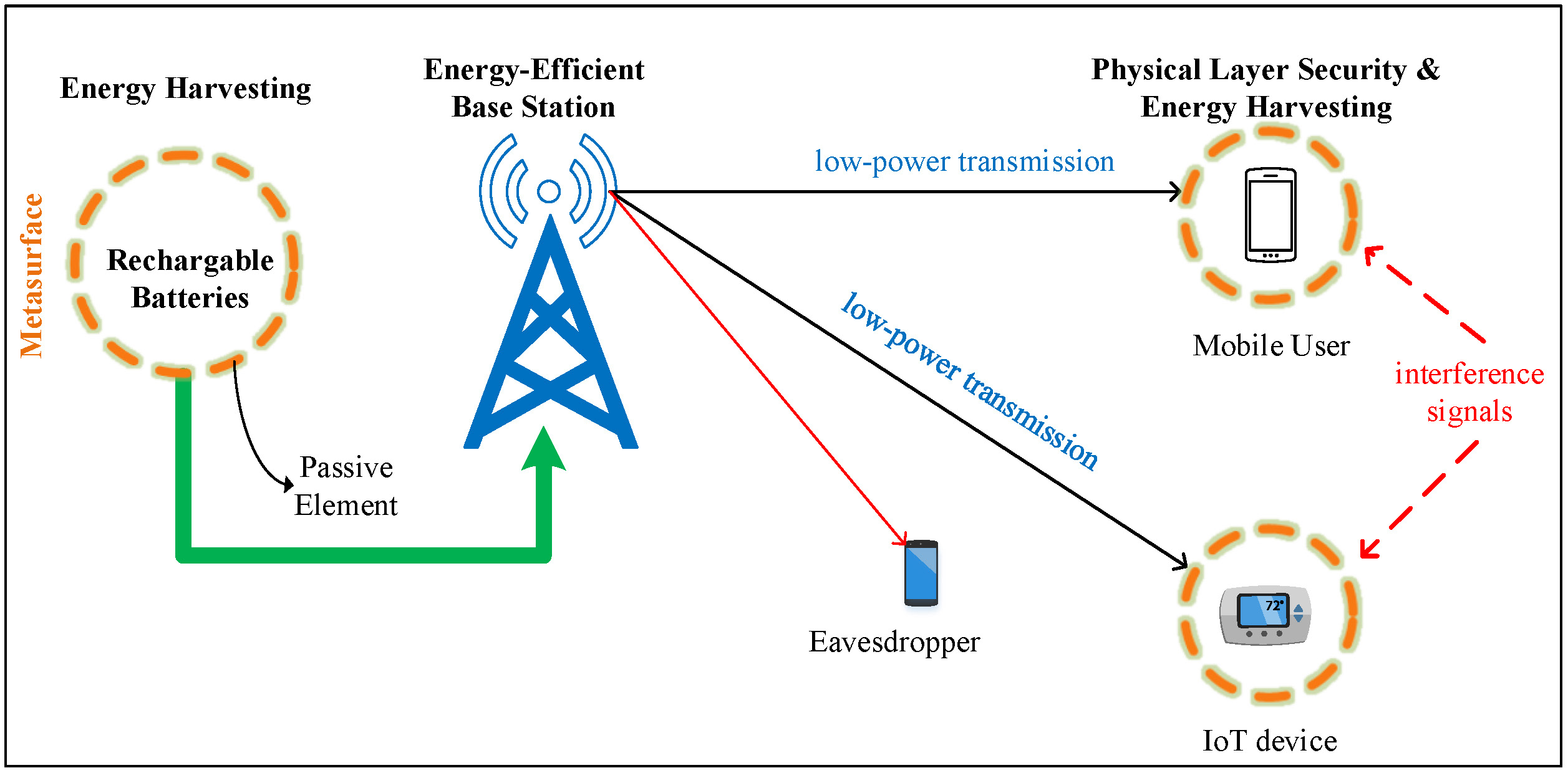}
	\caption{New paradigm: Metasurface coatings used for EE, EH and PLS for low-power transmissions.}\label{concept}
	\hrulefill
\end{figure*}
Prior to the 5G release, the technology was focused on improving the performance of communication systems in terms of error rate, outage and data throughput, assuming an unconfigurable propagation environment.  As a consequence, the main focus of the wireless society has been to utilize complex and inevitably energy-hungry transmission/reception as well as signal processing schemes capable of mitigating the channel impairments by exploiting new degrees of freedom. In particular, several utilities involve massive multiple-input multiple-output technologies, power allocation and coding through non-orthogonal multiple access, and new frequency resources in unlicensed bands, such as millimeter wave and terahertz, or underutilized ones by means of cognitive radio. Consequently, the energy sustainability of 5G networks becomes questionable, despite the efficiency of new power allocation algorithms and novel resource management schemes \cite{OULU6G}. 

Inspired by this, energy efficiency (EE) and autonomy of wireless devices have been identified as fundamental performance indicators of the next-generation networks. Likewise, security in wireless sixth-generation (6G) communications remains a critical aspect. Oppositely to encryption-based security methods, physical layer security (PLS) has appeared in the last decade to be a reliable alternative. PLS techniques take advantage of the physical characteristics of wireless channel in order to increase the date rate of the base station (BS)-legitimate user channel over the BS-eavesdropper channel. However, to guarantee security in wireless networks, more energy is consumed by transmitting artificial noise (AN) signals and utilizing secure beamforming or precoding schemes at the BS, being in contrast with ``green philosophy'' \cite{SecureLow-SNR}. In this sense, the EE-security trade-off appears to be hard in balancing; thus, it is still an open issue, having attracted a considerable amount of research interest \cite{SecureIoT}. To break the power consumption floor set by the 5G technologies and increase the energy autonomy of the wireless devices, novel energy harvesting (EH) approaches capable of absorbing energy from various sources, even from the interfering signals, were examined \cite{DesignEH}.

On another front, since metamaterials mimic exotic electromagnetic properties non-findable in nature, metasurfaces \cite{SmithReview} emulate exotic boundary conditions admitting general beam steering and field transformation. Indeed, placing multiple similar objects across a surface, can change the characteristics (phase, amplitude and polarization) of the waves passing through it, in an arbitrary way, and devise a favorable electromagnetic environment \cite{CapassoReview}. These applications and functionalities cover a broad range of operational wavelengths from microwaves to visible and may also concern non-electromagnetic interactions like acoustic ones. Introducing phase discontinuities within the thin parallelepiped volume of a metasurface separating two media, admitted the re-formulation of the laws of reflection and refraction by applying Fermat's principle \cite{GeneralizedLawsReflection} and revealed unexploited opportunities for light manipulation with planar photonics \cite{PlanarPhotonics}. 


Similarly, steering of wireless signals along various directions of space and with controllable magnitudes may be vital in overcoming physical obstacles in high-frequency communications, mitigating interference, and diminishing estimation error by comparing multiple weighted copies of the same waveforms. Additionally, passive metasurfaces can be efficiently employed as energy harvesters scavenging power from various interference and noise sources \cite{PadillaAbsorber}. It should be stressed that the fabrication of telecom-oriented metasurfaces is much easier compared to the construction of setups operated at shrunk wavelengths since printed-circuit approaches can be quickly and reliably adopted.

The software-defined reprogrammable metasurfaces that are administrated by at least one microcontroller are called, \textit{reconfigurable intelligent surfaces (RISs),} and have very recently begun to attract interest from the wireless communications engineering standpoint. Particularly, in \cite{MDRenzo_Metasurface}, the \textit{smart radio environments} concept has been introduced showing the great potential of RIS to manipulate the wireless media by changing specific characteristics of the signal. From a wireless communications theoretic viewpoint, the authors in \cite{Basar} have studied and compared the conventional communication models with the RIS-based one, while in \cite{Valagiannopoulos_2019}, a simple nanoslit metasurface radically increases the directivity of the transmitting beams leading to significant enhancement of the signal-to-interference ratio (SIR) in a visible light communication system. Moreover, two promising programmable metasurface-based RF-chain designs that aim at reducing the hardware cost has been recently presented in \cite{RIS_Transceiver}. In the important work \cite{Huang_TWC}, the authors developed a RIS power consumption model that is based on the number of deployed reflector elements and their phase resolution capability while two novel low complexity optimization algorithms were also developed for EE maximization. 


Although RIS technology has been identified as a main pillar of 6G systems, there are still several practical issues that need to be resolved. In more detail, channel estimation problems, phase shift inaccuracies as well as reconfiguration of metasurface elements in real time, arise. Additionally, the EE, EH, and PLS appear not to be fully controllable by RIS considering that the maximization of the signal-to-noise ratio (SNR) at the destination node is conditioned on the reflected angle. Therefore, a new paradigm for efficient utilization of passive metasurfaces to jointly address the problems stated above is~necessary.

In this paper, we tackle three critical issues of ultra-high EE transmission, practical EH devices, and PLS for ultra-low-power (ULP) wireless transmissions in real environments. Contrarily to the conventional mode of operation of RIS, where in the vast majority of application scenarios used as reflectors between transmitter and receiver with reflecting elements adjusted by a microcontroller, the proposed novel \textit {power-collecting metasurface coats } are respectively deployed at either the transmitter or receiver of the wireless system under consideration. Consequently, the power-collecting functionalities of metasurfaces are applied to realize the ULP transmissions, ultra-high EE, EH, and PLS. The novelties and technical contributions of the proposed paradigm are threefold: 
\begin{itemize}
	\item \textit{Power allocation algorithms in ULP access points (APs):} Without employing complicated phase control routines and excessive channel estimation, as in conventional RIS-based solutions, the proposed power allocation algorithms can meet the key performance requirements and ensure high-quality of experience (QoE) for new scenarios under ULP transmissions.
	\item	\textit{Metasurface-coated devices:} The use of passive-metasurface as a shell not only absorbs the power of RF signals in the ambient environment, but also reduces the level of interference. Therefore, the destination terminal can detect the power of the source signal, even if it is transmitted at ULP levels. After wrapping the user with the proposed power-concentrating metasurface, the device can perform EH from all kinds of RF signals developed nearby; hence, the interference becomes one of the available energy sources for improving EH.
	\item	\textit{Metasurface-enabled PLS:} The metasurface-coated devices proposed in this paper are applied to improve the PLS performance in the sense of EH and improved  power levels at the receiver side. The latter realizes ultra-high-power collection considering the trade-off between EH and PLS. 
\end{itemize}

The rest of this paper is organized as follows: In Section II, we describe the new concept of metasurface claddings and in Section III we present the EM theory on metasurface-covered enabled communication with application to EE, EH and security; the significance of the proposed concept is highlighted through preliminary simulation graphs. Based on these results, we then discuss challenges and research directions in Section IV before formulating concluding remarks in Section V.

\begin{figure}
	\centering
	\includegraphics[keepaspectratio,width=3.3in]{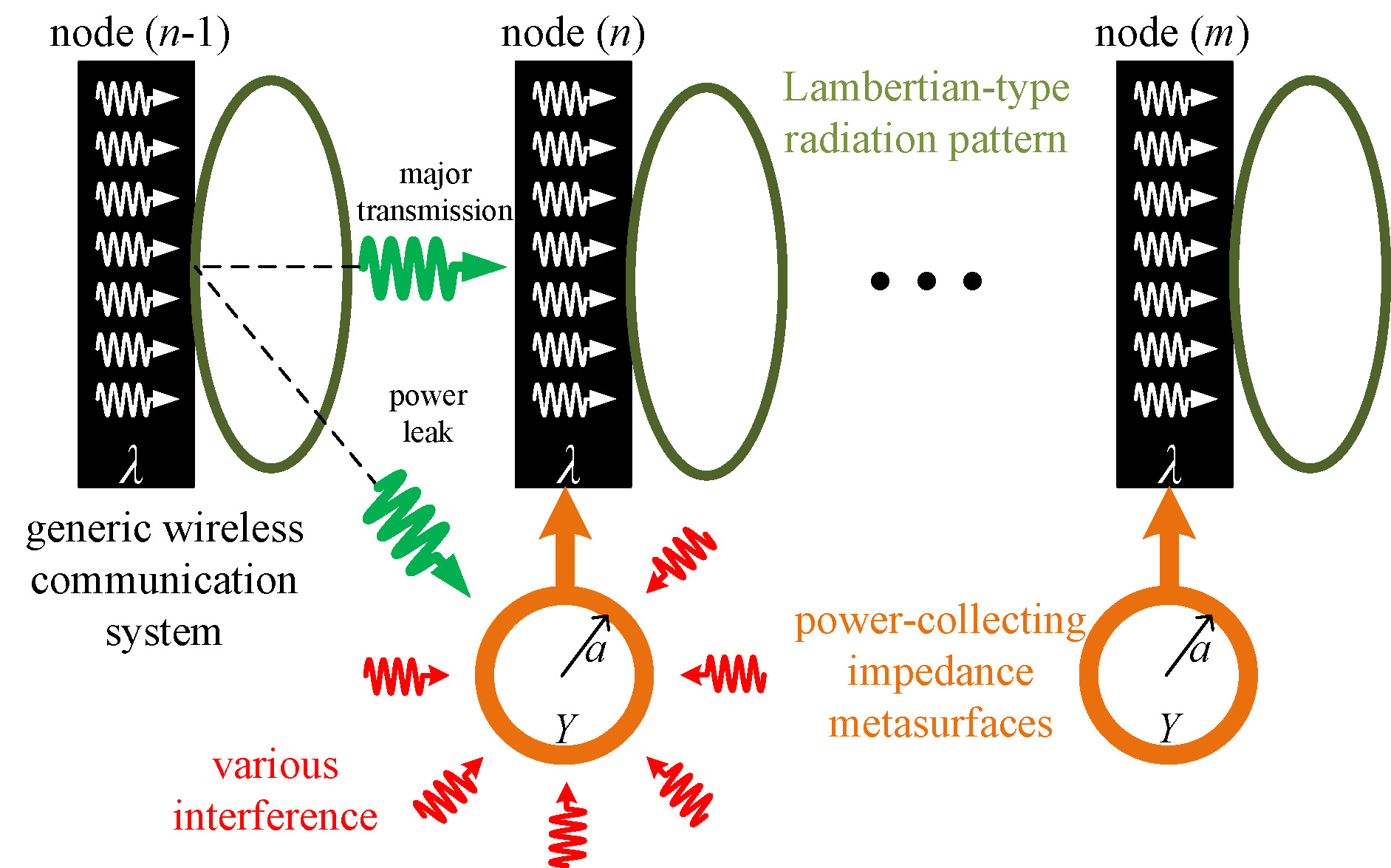}
	\caption{The proposed concept with application to EH communication system working at wavelengths $\lambda$. Every node of communication network with serial number $n$ is fed by the previous one with serial number $(n-1)$ and assisted by an impedance metasurface responsible for partially feeding it with power concentrated into its volume. The system is in chain connection and can be extended to an arbitrary node of serial number $m$.}
	\label{fig1}
\end{figure}

\section{Metasurface-Coated Devices: The Concept}
Fig.~\ref{concept} illustrates a novel paradigm of metasurface shells able to increase the EE, EH and PLS. The utilization of metasurface-coated devices is conceptually different from the conventional RIS-based ones. In principle, green communication is implemented at the BS by using EH and ULP transmissions, since the devices harvest energy while being highly secured from unwanted interception. This holistic approach can be achieved by the unique electromagnetic properties of a metasurface, which further amplify energy reservations, reduce significantly the transmit power as well as hugely increase the secrecy rate. The metasurfaces can be installed around small-size devices (e.g., IoT machines, sensors, mobile phones, etc.) and potentially incorporate a pack of rechargable batteries to cover the energy demands of small-cell APs.

To further elaborate on the differences between the two concepts, RIS is a nearly passive solution since a microcontroller is needed to adjust the phases of the incident wireless signals in a deliberate manner to maximize the SNR at the destination terminal. However, the configuration of the optimal phase shifts of RIS meta-elements is very challenging since the channel state information (CSI) of the cascaded RIS channel (Tx-RIS-Rx) cannot be easily acquired due to the passive nature of RIS. By assuming ideal CSI, which is the case in most publications, RIS-enabled wireless communications can be a viable solution and improve the overall performance of the wireless network. Nevertheless, our proposed solution is conceptually different and less complicated than RIS, i.e., totally passive without a microcontroller. That is, the fundamental difference between the conventional RIS and our proposed metasurface-coated device is \emph{reconfigurability}. With the absence of reconfigurability, the proposed device can reduce the operational complexity compared to the RIS-aided systems, while losing some degrees of freedom in maximizing the power collection. Furthermore, the proposed passive metasurface coats address three challenging issues:  EE, EH, and PLS in ULP wireless transmissions. In opposition to the operational mode of RIS, the proposed metasurface claddings are deployed either at the transmitter (pure EH mode) or receiver ends of a wireless communication system aiming at collecting the maximum signal power available in the surrounding environment. Interestingly though, one can design the described setup by taking into account the power distribution over frequency in the vicinity of the metasurface, and selecting proper materials whose dispersion allows for wideband operation.

\textbf{Remark:} At this point, we have to emphasize that, although the current solution significantly improves the power concentration, the SNR at each user may not directly be enhanced. On the one hand, since the desired signal collected power increases, the SNR should be improved. However, as the received power grows larger, according to the Bussgang theorem, the distortion noise that comes from non-linear sources, such as the receiver's low-noise amplifiers (LNAs), linearly increases, if and only if LNAs operate beyond the saturation point. This motivates the formulation of optimization problems that returns design guidelines for the LNA operation in metasurface-coated devices. Moreover, it is worth noting that the power-collecting metasurface coating combats the distance path-loss and large-scale fading, which should be considered in highly-dense urban areas. Consequently, the ultra-high-power concentration can efficiently balance the shadowing effect and significantly improve the outage performance by increasing the service area~efficiency. Finally, metasurface-coated devices manipulate the nearby electromagnetic environment; as a result the levels of co-channel interference, due to transmission of neighbor nodes, do not solitarily depend on the distance between the desired user and the interferers, but also on their angular position. Accordingly, the SIR is a function of the desired transmitter and interferers' powers, their distances, as well as their orientation with respect to the desired user.

\section{EM Theory for Metasurface-Coated Enabled Communication}
\subsection{Application to Energy Harvesting}

In Fig.~\ref{fig1}, we depict the operation of closed metasurfaces characterized by a homogenized complex admittance $Y$ used for EH assisting various nodes of an integrated communication system. In particular, we show a relay network with multiple nodes where a wireless emitter of serial number $(n-1)$ transmits directionally towards the receiver with serial number $n$. However, due to existence of secondary lobes, it inevitably loses power towards other directions; such a power leak can be collected, together with interference, by the proposed impedance metasurface of closed shape installed next to another node (with serial number $n$). Thus, a significant part of the energy required for the propagation of information signal is provided by other random signals developed in the vicinity of the transmitter being connected with the aforementioned power-collecting metasurface. This concept can be applied to each one of the involved nodes (transmitters, repeaters, and receivers) leading to a multiple and simultaneous implementation; in Fig.~\ref{fig1}, we sketch the network up to the node with serial number $m>n$. The radiation pattern of each emitter is assumed to be of Lambertian type, namely is maximal along the normal direction to the flat surface and vanishes at the tangential directions. 

\begin{figure}
	\centering
	\includegraphics[  keepaspectratio, width=1.7in]{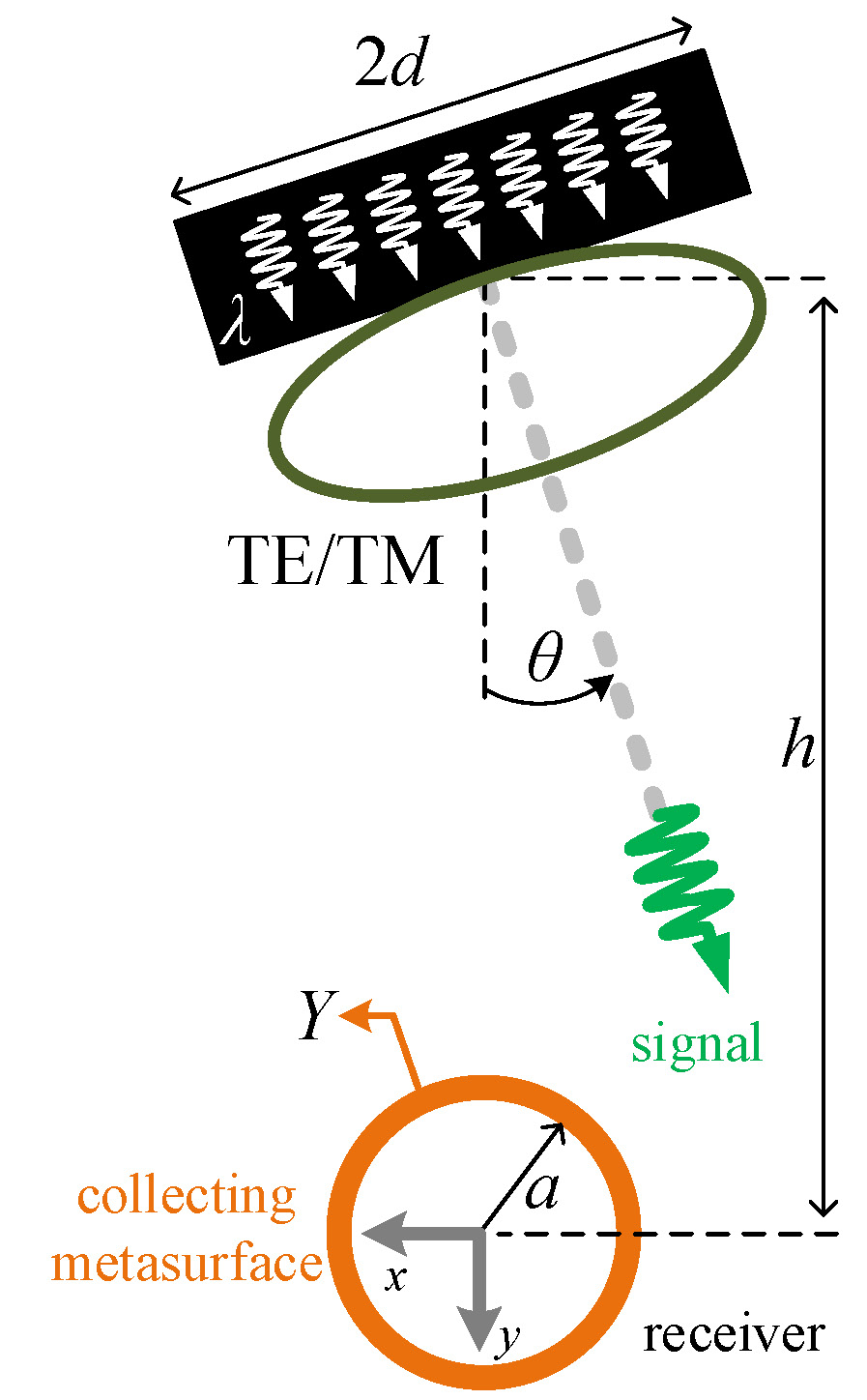}  	
	\caption{Sketch of the transmitter/receiver setup – the source is a radiating aperture with dipole-like radiation pattern. }	\label{fig3}
\end{figure}

The behavior of our homogenized metasurface can be characterized through the real and imaginary parts of its surface admittance $Y$. According to Ampere’s law $\nabla\times\mathbf{H}=\mathbf{J}+\frac{\partial \mathbf{D}}{\partial t}$
and the microscopic Ohm’s law across a surface $\mathbf{J}_s=Y\mathbf{E}$,
the real part of the surface admittance $\textrm{Re}[Y]$ determines if the structure pumps or absorbs energy to the system; thus, it is labeled as active or passive respectively. On the other hand, the imaginary part $\textrm{Im}[Y]$ concerns the phase difference in harmonic response and gives dielectric or plasmonic character to the wave interactions. Obviously, a vanishing real part $\textrm{Re}[Y]=0$ describes lossless designs. This issue can be further explained if one takes into account that the surface admittance $Y$ is mimicked by a thin slab of dielectric permittivity $\varepsilon$ with $\varepsilon-1=\frac{Y\eta_0}{i k_0g}$ where $g$ and $\eta_0$ stand for the slab thickness and wave impedance in free space, respectively. In this way, the polarization vector $\textbf{P}$ is defined as $\textbf{P}=(\varepsilon-1)\textbf{E}=\frac{Y\eta_0}{i k_0g}\textbf{E}$. It is clear that if $\textrm{Im}[Y]>0$, the two vectors $(\textbf{P},\textbf{E})$ are pointing to the same direction which indicates the dieletric behavior; on the contrary, if $\textrm{Im}[Y]<0$, the polarization $\textbf{P}$ is opposite to \textbf{E} and the nature of the system is plasmonic.

In Fig.~\ref{fig3}, we show an indicative setup of a misaligned transmitter-receiver (by angle $\theta$), where the discussed metasurface is used to wrap the latter one. The source is a radiative aperture of size $2d$ working close to RF wavelength $\lambda$ and possesses a dipole-like radiation pattern. The polarization of the emitted signal can be of both types (transverse magnetic (TM) and transverse electric (TE)) and the distance between the transmitter and the collector is denoted by $h$. The power concentration area is defined by the considered cylindrical metasurface of radius $a$ and surface admittance $Y$ into which our actual receiving system is placed.

\begin{figure*}
	\begin{subfigure}{.4\textwidth}
		\centering
		\includegraphics[keepaspectratio, width=3.5in]{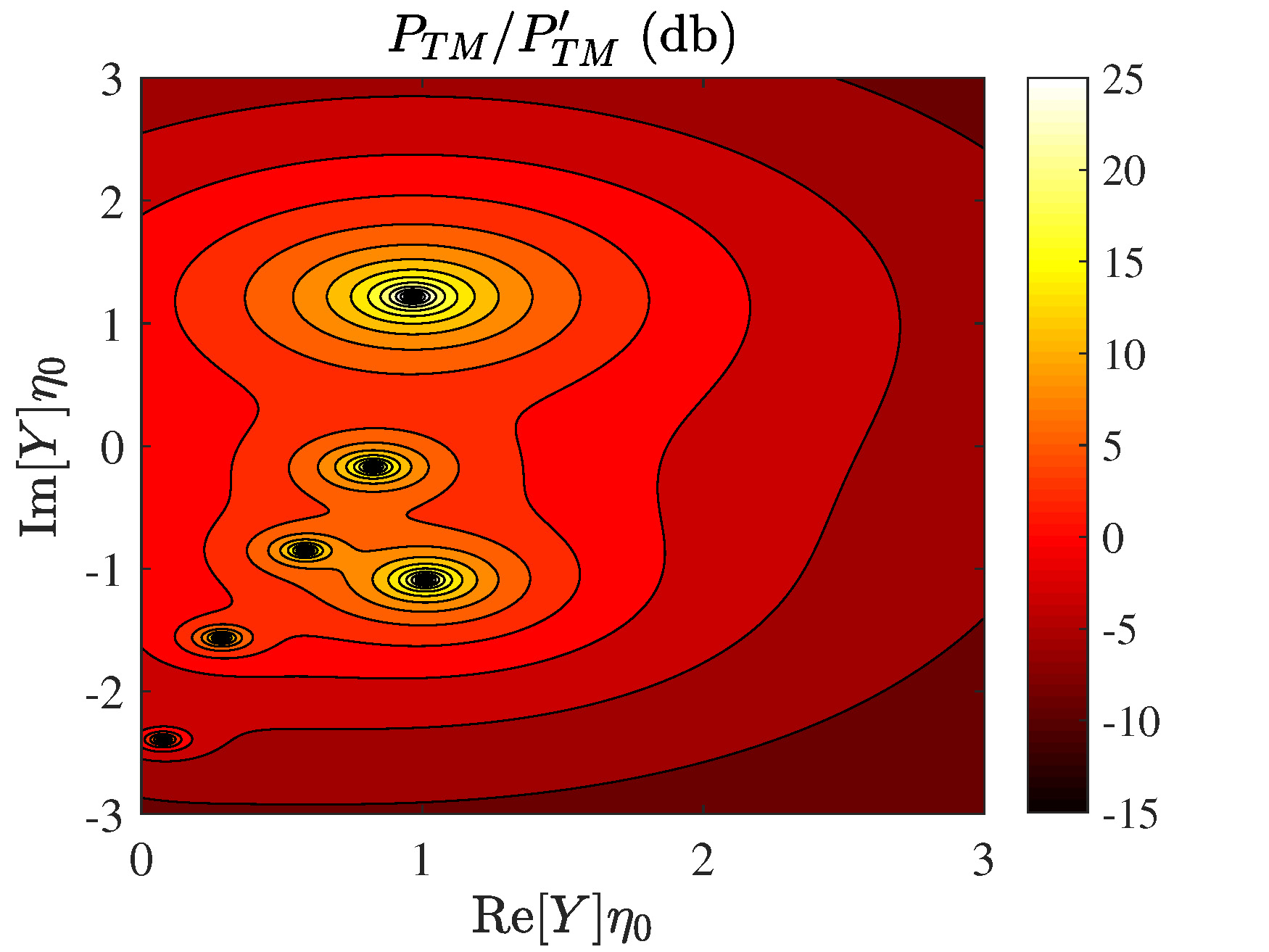}  
		\caption{}
		\label{fig4a}
	\end{subfigure}
	\hfil
	\begin{subfigure}{.4\textwidth}
		\centering
		\includegraphics[keepaspectratio, width=3.5in]{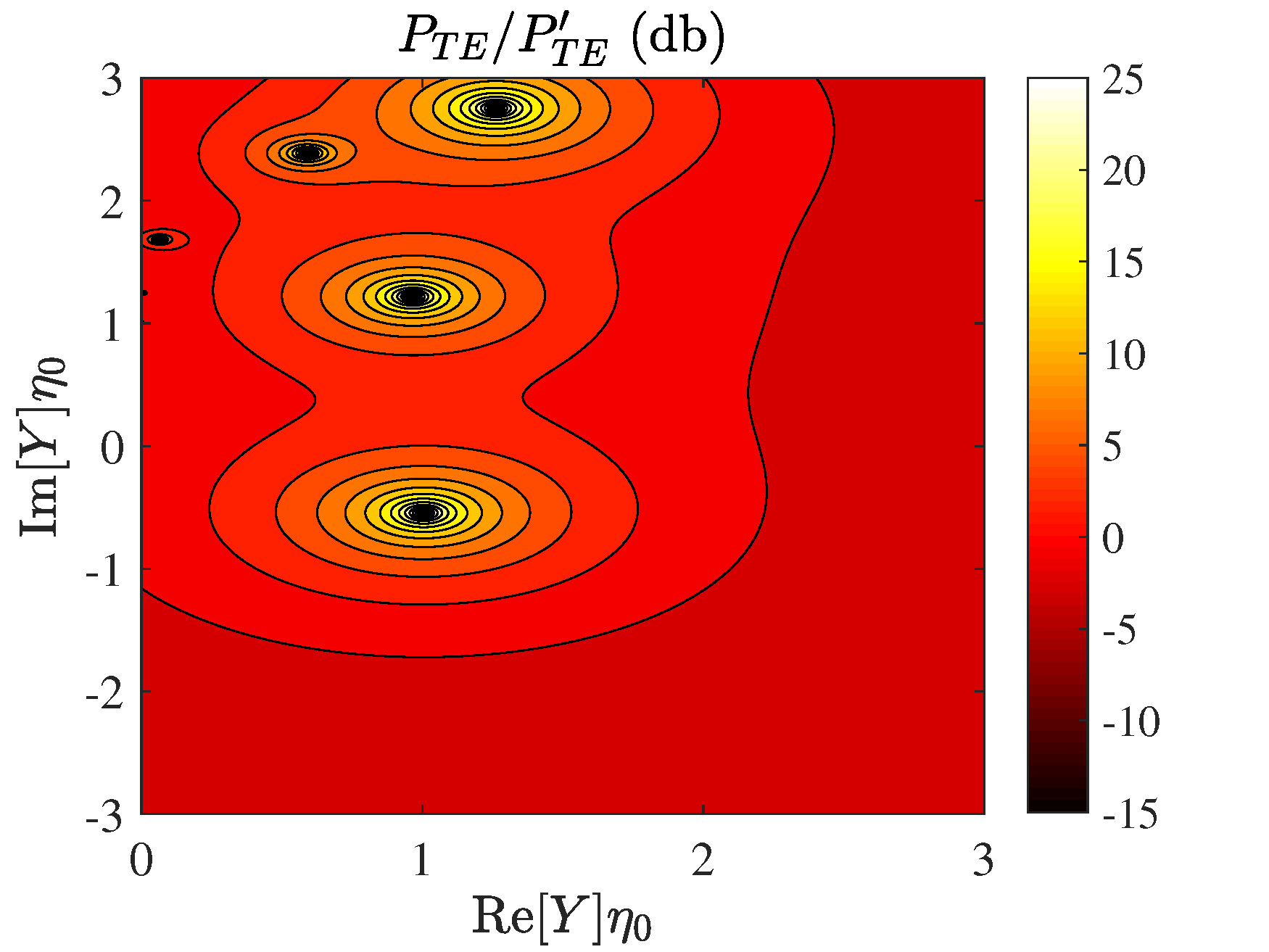}  
		\caption{}
		\label{fig4b}
	\end{subfigure}
	\caption{Enhancement of concentrated power into the circle of radius $a$ on the complex plane of surface admittance $Y$ for (a) TM and (b) TE waves. }
	\hrulefill
\end{figure*}

To quantify the EH performance of the proposed architecture in Fig.~\ref{fig4a}, we consider a metasurface of circular shape with radius $a=5$ cm illuminated by an RF dipole source operated at $\lambda=10$ cm, i.e., $300$ GHz, positioned at distance  $h=100$ cm; the boundary value problem is solved with rigorous electromagnetic formulation. The represented quantity is the ratio of the signal power across the circle’s area after ($P$) over the same quantity before ($P^\prime$) the metasurface installation; all the values onto the complex plane of surface admittance $Y$ are swept. For example, in the TM scenario, namely when the electric field is normal to the two-dimensional plane of Fig.~\ref{fig3}, it is written inside the metasurface as follows:
\begin{align}
	E_{z,rec}=\sum_{n=-\infty}^{+\infty}{A_nJ_n\left(k_0r\right)e^{in\varphi}}
	\label{Ezrec}
\end{align}
where $\left(r,\varphi\right)$ are the local polar coordinates, $k_0=2\pi/\lambda$ is the wavenumber into free space, $J_n$ the Bessel function of $n$-th order and $A_n$ complex determinable constants via the imposed boundary condition:
\begin{align}
	\hat{\mathbf{r}}\times\left[\left.\mathbf{H}_{out}\right|_{r=a}-\left.\mathbf{H}_{rec}\right|_{r=a}\right]
	=-Y \hat{\mathbf{r}}\times\left[\hat{\mathbf{r}}\times\left.\mathbf{E}_{out}\right|_{r=a}\right],
\end{align}
where $\hat{\mathbf{r}}$ is the radial unitary vector being normal to the boundary of our metasurface. The notation $\{\mathbf{E}_{out},\mathbf{H}_{out}\}$ is used to denote the electromagnetic field out of the metasurface ($r>a$) while the symbols $\{\mathbf{E}_{rec},\mathbf{H}_{rec}\}$ stand for the received electromagnetic field in the cylinder ($r<a$). The electric components that are tangential to the cylinder are continuous around the circular boundary $r=a$, namely $\hat{\mathbf{r}}\times\left.\mathbf{E}_{out}\right|_{r=a}=\hat{\mathbf{r}}\times\left.\mathbf{E}_{rec}\right|_{r=a}$. As far as the tangential magnetic components are concerned at $r=a$, they are discontinuous by a quantity proportional to the surface admittance $Y$ and the common tangential electric field at the boundary. The power per axial unit meter of the signal developed across the cavity, takes the following rigorous form:

\begin{align}
	P=\frac{2\pi a}{\eta_0}\sum_{n=-\infty}^{+\infty}{\left|A_n\right|^2\left[J_n\left(k_0r\right)\right]^2}. 
	\label{Prec}          		
\end{align}
In Fig.~\ref{fig4a} the considered polarization is of TM type (electric field $\mathbf{E}$ normal to the plane of Fig.~\ref{fig3}) and in Fig.~\ref{fig4b} the waves are of TE nature (magnetic field $\mathbf{H}$ normal to the plane of Fig.~\ref{fig3}). One directly notices that in both cases the power enhancement can be of 2-3 orders of magnitude; thus, the beneficial influence of the power-collecting metasurface is soundly indicated. It is also remarkable that there are regions of the parametric space $\left(\textrm{Re}\left[Y\right],\ \textrm{Im}\left[Y\right]\right)$ where the power enhancement both TM $\left(P_{TM}/P_{TM}^\prime\right)$ and TE $\left(P_{TE}/P_{TE}^\prime\right)$ polarizations reaches substantial levels.
\begin{figure*}
	\begin{subfigure}{.3\textwidth}
		\centering
		\includegraphics[width=2in]{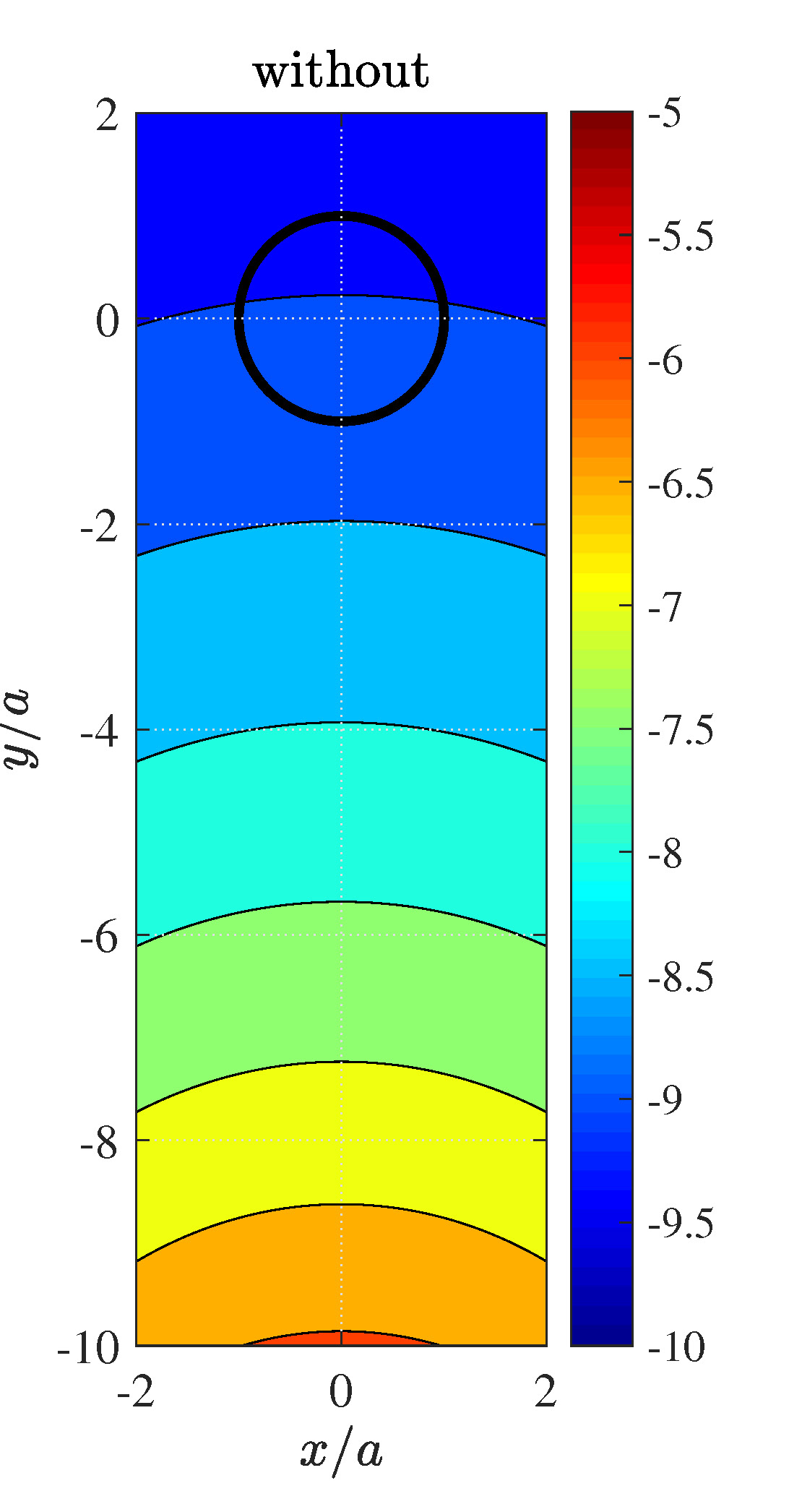}  
		\caption{}
		\label{fig5a}
	\end{subfigure}
	\begin{subfigure}{.37\textwidth}
		\centering
		\includegraphics[ width=2in]{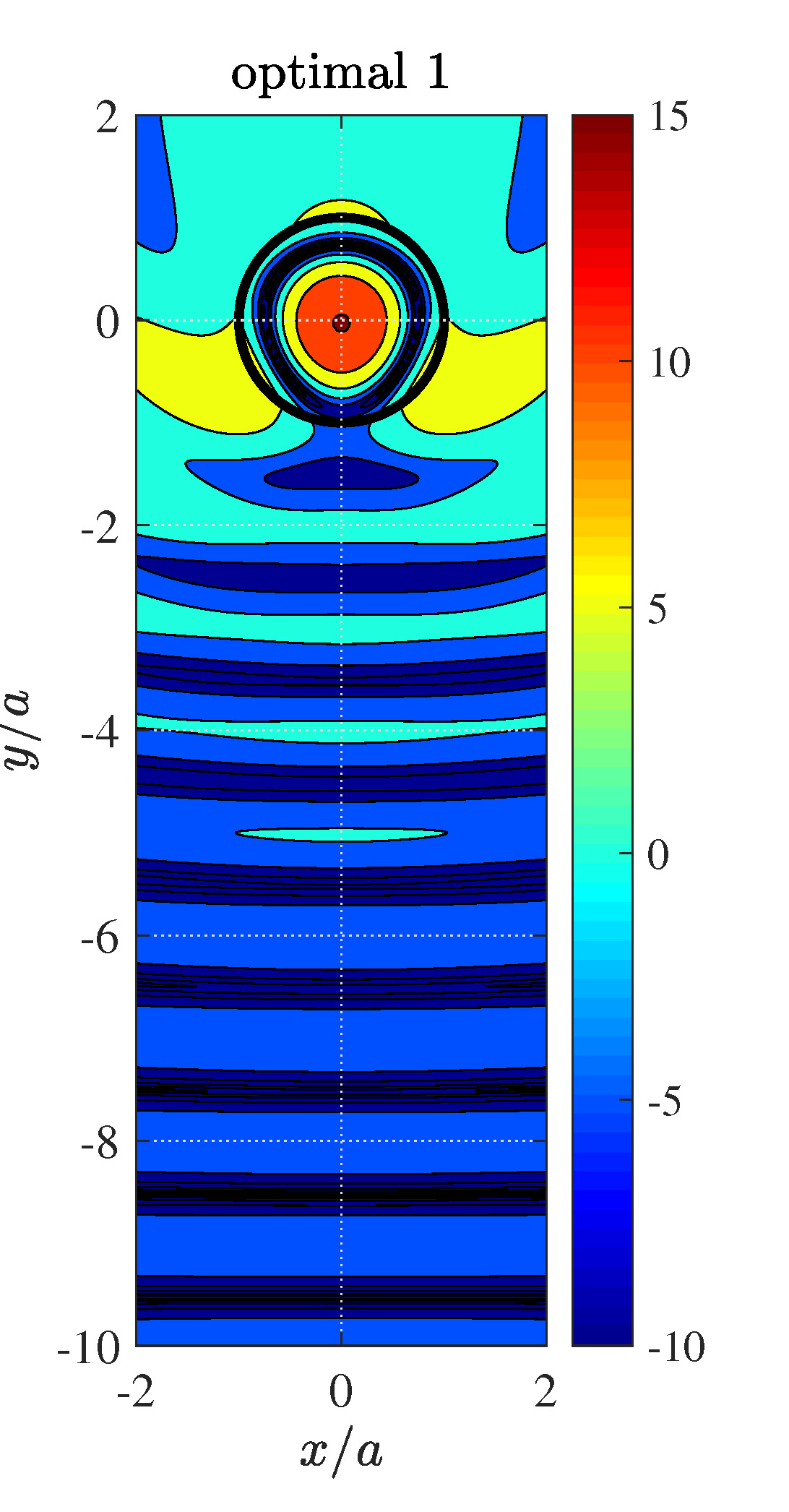}  
		\caption{}
		\label{fig5b}
	\end{subfigure}
	\hfil
	\begin{subfigure}{.3\textwidth}
		\centering
		\includegraphics[ width=2in]{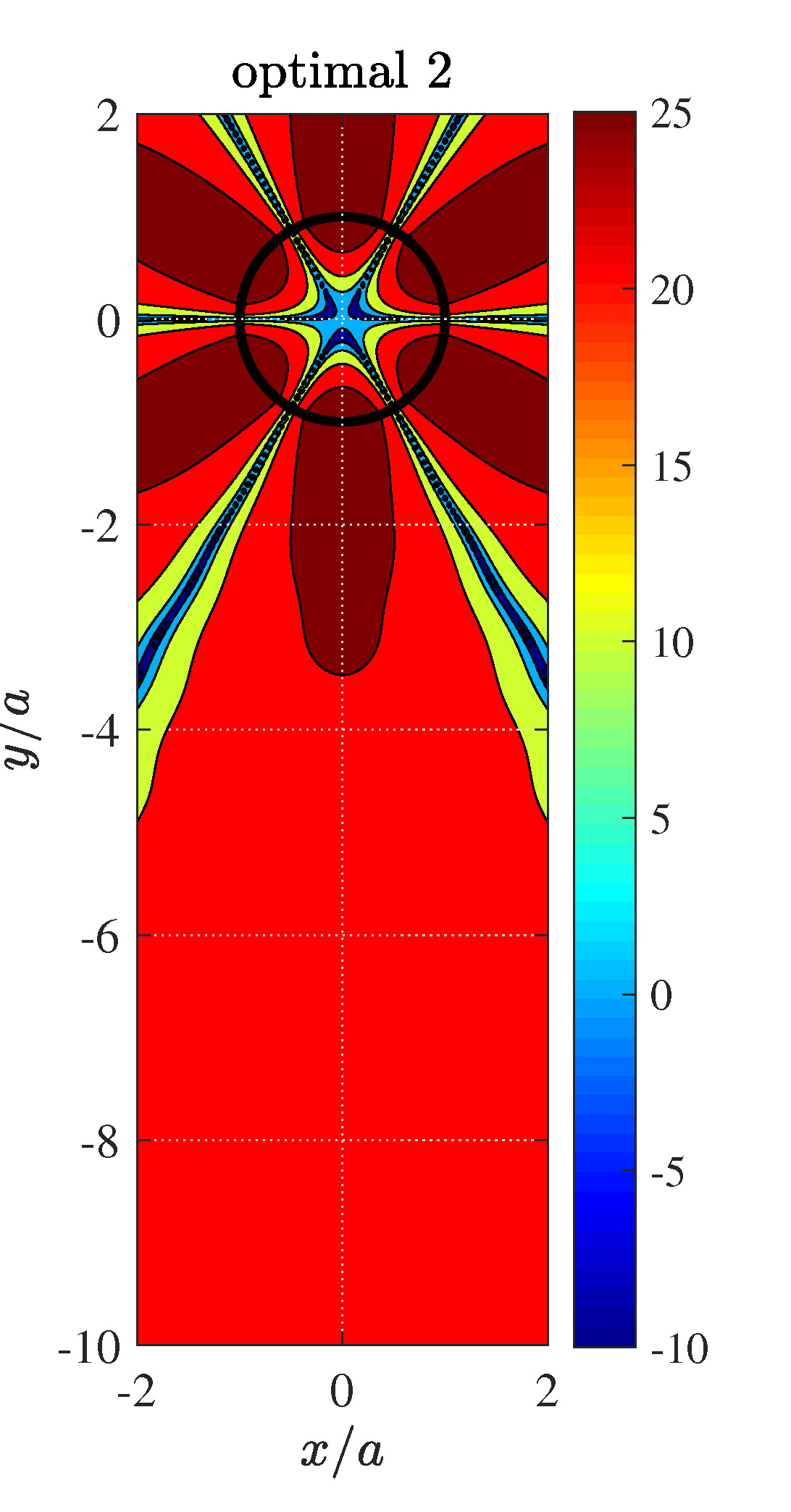}  
		\caption{}
		\label{fig5c}
	\end{subfigure}
	\caption{Spatial distribution of the squared value of the electric field magnitude in dB for: absent metasurface (a), one optimal metasurface (b) and another optimal metasurface (c). Black loop indicates the position of the circular metasurface, even though it has been removed as in (a).}
	\label{fig5}
	\hrulefill
\end{figure*}
For the sake of comparison with conventional approaches, in Fig.~\ref{fig5} shown at the top of the next page, we demonstrate the spatial distribution of the electric field $\left|E_z\left(x,y\right)\right|^2$ when no metasurface is deployed (Fig.~\ref{fig5a}) and if one uses two (Figs.~\ref{fig5b} and \ref{fig5c}) of the optimal metasurfaces indicated by Fig.~\ref{fig4a}. In particular, the Fig.~\ref{fig5b} corresponds to a maximum enhancement of Fig.~\ref{fig4a} giving an omni-directional resonance for the cylindrical cavity defined by the considered metasurface. We notice a large boost of the average signal power into the metasurface surpassing the two orders of magnitude. The enhancement is even more impressive in the Fig.~\ref{fig5c} where the sustained resonance is of higher angular momentum order and several lobes are formulated around the cylinder. The order of the sustained or activated resonance is the number $n$ from the equation of electric field at the receiver, showing which term of the sum  is maximized. In this way, we obtain a strong response with azimuthal profile $e^{i n \varphi}$ from \eqref{Ezrec}, which corresponds to larger variation with respect to angle $\varphi$, namely higher angular momentum order. This indicates that if one is interested about the average signal across the cylinder and not the spatial distribution of the power in its vicinity, the excitation of higher order resonances is preferable. In Fig.~5b, we have a central hotpoint, while in Fig.~5c the signal concentration is more substantial but extends also outside of our metasurface. It should be stressed that every single of the maxima of Figs~\ref{fig4a} and \ref{fig4b} correspond to spatial distributions like them in Figs~\ref{fig5b} and \ref{fig5c} revealing the existence of several resonances.

To put it alternatively, we are in search of resonances, namely of maximization of the sum \eqref{Ezrec}. This routinely happens when one of the terms of the series increases substantially and becomes dominant over the others. The number $n$ corresponding to that term gives the order of the resonance. Note that each term comprises of a $\varphi$-dependence given by $e^{i n \varphi}$ meaning that the larger $n$ is, the more substantial is the $\varphi$ variation of the signal patterns. The distributions of Figs 5b and 5c correspond to different maxima of Fig. 4a, namely resonances of different orders $n$. It is demonstrated by inspection of Fig. 5b which indicates that $n=0$ since the field concentration is not significantly dependent on angle $\varphi$. On the contrary, Fig. 5c reveals the presence of a higher angular momentum resonance with $n=3$ since the represented quantity is the squared magnitude of the electric field.

\subsection{Application to PLS}

The operation of power-collecting metasurfaces for PLS can be described as follows: If the transmitter emits very low power, below the sensitivity of scattered eavesdroppers, it is practically invisible to them. However, the receiver that employs the proposed metasurface enhances the signal power admitting it to properly record the information. In particular, the circular metasurface deforms locally the propagating waves and attracts them along all possible directions around it. It should be stressed that, in this simplistic scenario, the SIR at the receiver is not enhanced, since interference gets amplified proportionally to the information signal. In this way, one can establish reliable communications by emitting through a pool of unfriendly receptors. The same metasurface can be used for the opposite purpose by wrapping a spying aperture, which is located far away from the major source. Therefore, the communication channel can be overheard by someone outside the vision range of the emitter.

\begin{figure*}
	\begin{subfigure}[b]{0.5\textwidth}
		\centering
		\includegraphics[ keepaspectratio, width=\columnwidth]{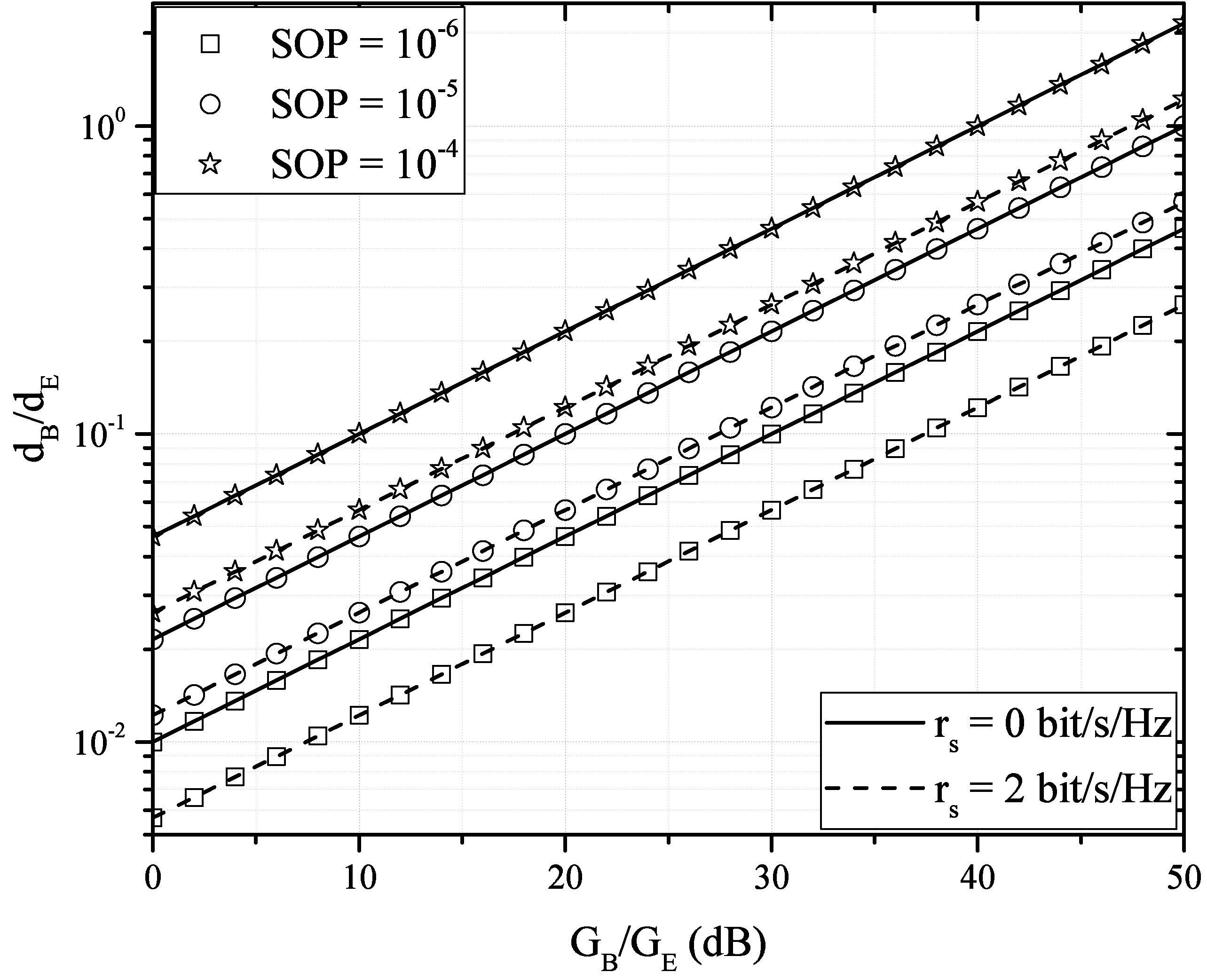}
		\caption{ }
	\end{subfigure}
	\begin{subfigure}[b]{0.56\textwidth}
		\centering
		\includegraphics[ keepaspectratio, width=\columnwidth]{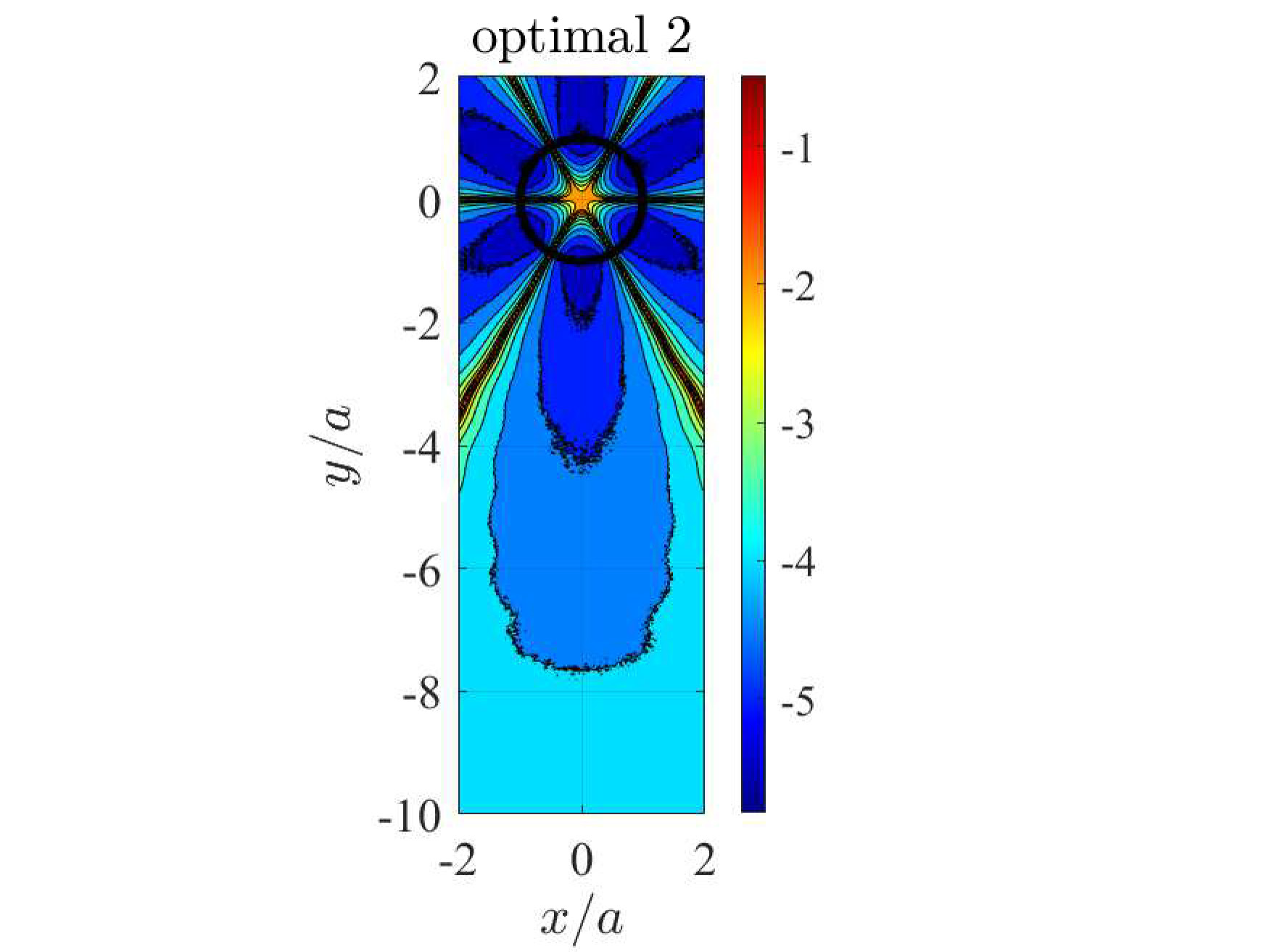}
		\caption{ }
	\end{subfigure}
	\caption{(a) Maximum allowed legitimate user to eavesdropper distances ratio vs legitimate user to eavesdropper coating gains ratio, for different values of SOP and $r_s$ requirements. (b) Spatial distribution of $\log\left(\text{SOP}\right)$ for $r_s=0$, assuming that the legitimate user is covered by the optimal 2 metasurface, while the eavesdropper uses no metasurface.}
	\label{fig7}
	\hrulefill
\end{figure*}

To quantify the metasurface-coated device gain in terms of PLS, Fig.~\ref{fig7}a illustrates the maximum allowed legitimate user to eavesdropper distance ratio as a function of the legitimate user to eavesdropper coating gain ratio, for different values of secrecy outage probability (SOP) and secrecy rate, $r_s$, requirements. SOP is an essential performance metric of security, which is defined as the probability that difference between the capacity of the main channel and capacity of the channel malicious intruder to become lower than a predetermined threshold. In other words, it defines the maximum rate at which data can be communicated from the transmitter to the legitimate receiver under the appropriate secrecy level. In Fig.~\ref{fig7}a, $G_B$ and $G_E$ respectively denote the gains due to metasurface coat of the legitimate user and eavesdropper.  Likewise, $d_B$ and $d_E$ respectively stand for the legitimate user and eavesdropper distances from the transmitter. In the scenario under investigation, it is assumed that both the legitimate user and eavesdropper are steered towards the transmitter in a way that the maximum gain can be achieved. Moreover, the Rayleigh distribution is used to model the small-scale fading in both the transmitter legitimate user and transmitter eavesdropper links. As expected, for fixed  $r_s$ and SOP requirement, as $\frac{G_B}{G_E}$ increases, the $\frac{d_B}{d_E}$ also increases. For example, for  $r_s=0$ and SOP equals $10^{-6}$, as $\frac{G_B}{G_E}$ increases from $0$ to $30\text{ }\mathrm{dB}$,  $\frac{d_B}{d_E}$ increases by one order of magnitude. Another interesting observation that can be extracted from this figure is that for specific $r_s$ and SOP requirements, always exists a $\frac{G_B}{G_E}$ threshold beyond which $\frac{d_{B}}{d_{E}}$ becomes greater than $1$. For example, by setting the SOP requirement to $10^{-4}$ and for $r_s=0\text{ }\mathrm{bit/s/Hz}$, $\frac{d_{B}}{d_E}$ is greater than $1$, for $\frac{G_{B}}{G_{E}}\geq 40\text{ }\mathrm{dB}$. This indicates that metasurface-coated device enables the relaxation of the PLS limitation according to which the legitimate user should be closer to the transmitter compared to the eavesdropper. Additionally, from this figure, we observe that, for given $\frac{G_B}{G_E}$ and SOP, as $r_s$ increases, the maximum allowed $\frac{d_{B}}{d_{E}}$ decreases. As an example, for $\frac{G_B}{G_E}=30\text{ }\mathrm{dB}$ and SOP equals $10^{-6}$, the maximum allowed $\frac{d_{B}}{d_{E}}$ decreases by approximately $10$ times, as $r_s$ increases from $0$ to $2\text{ }\mathrm{bit/s/Hz}$. Finally, for fixed $\frac{G_B}{G_E}$ and $r_s$, as the SOP requirement relaxes, the maximum allowed $\frac{d_{B}}{d_{E}}$ increases.  

Fig.~\ref{fig7}b illustrates the spatial distribution of $\log_{10}\left(\text{SOP}\right)$, for $r_s=0$. In this scenario, it is assumed that the legitimate user is placed in $(0, 0)$ and is covered by the optimal 2 metasurface, used also in~Fig.~\ref{fig5c}. On the other hand, the eavesdropper uses an omni-directional antenna, like the one depicted in Fig.~\ref{fig5a}. We assume that the eavesdropper can be placed in any location. Finally, the transmitter is placed in a point on $x/\alpha=0$ and $y/\alpha<0$, such that the received power at the legitimate user is $10\text{ }\mathrm{dBm}$. From this figure, it becomes apparent that there are only specific angular positions of relatively short length in which the SOP does not surpass $10^{-4}$. This indicates that metasurface-coated devices can become a fundamental pillar of a next-generation of physical layer security schemes.
\section{Challenges \& Research Directions}
This section is devoted on presenting possible research directions on the topic of metasurface cladding. 

\subsection{Ultra-High EE in ULP Transmission} 
The critical problem, which is addressed with the novel power-concentrating metasurfaces is the unprecedented reduction of the amount of energy transmitted from BS to all devices. Now, with the proposed versatile metasurface covers, not only the energy consumption is significantly reduced at the BS due to ULP transmissions, but also the energy collected at the receiver side is enormous, and the wireless links are more secured than ever before. The devices can leverage both the electromagnetic radiation from multiple known and unknown transmissions for harvesting purposes and, at the same time, drastically reduce the risk of eavesdropping. In this context, metasurface claddings shape a new approach for energy management and PLS in 6G networks.

\subsection{Practical EH Based on Metasurface Coats}
The current EH approaches, in wireless systems, refer to how much power are able to collect. The passive metasurface wrapping will provide a substantial increase in the power that can be gathered from various signal transmissions; accordingly, the power-collecting metasurface will address a critical problem of the energy that can be scavenged wirelessly. The proposed impedance coat is capable of providing (theoretically) up to 25 dB power enhancement and seems to be a very promising method for recharging not only small devices, but equipment with much more significant needs for energy, e.g., BSs for small cells.

\subsection{Security in ULP Profile Transmitters} The critical problem of securing wireless communications links in ULP transmission can be studied in future. Up to now, all the solutions are mostly based on the transmitter by using beamforming, jamming techniques, and AN transmissions in order to secure confidential messages from overhearing. However, the metasurface coating transfers the security capabilities at the receiver side by using a passive metasurface that is capable of increasing the gathered power significantly even at great distances from the transmitter or in a short distances in a ultra-low source transmit power profile. So, the critical problem of the trade-off between ULP transmissions and secrecy capacity is addressed for the first time and can be the foundation for future research.

\subsection{Power Conversion Models}
The current energy/power conversion models existing for standard wireless power transmission are no longer applicable; instead, new power conversion efficiency models based on power, frequency and distance should be formulated. After adopting the metasurface cladding concept, the radio propagation and signal transmission/reception models under ULP transmissions framework become essentially different from the conventional ones. EH should be designed through advanced electromagnetic field modeling and electromagnetic property testing; therefore, new metasurface-enabled power conversion models should be proposed and analytically studied under ULP-profile BSs.

\subsection{A New Electromagnetic-Theory-Driven Information Theoretic Framework that Accommodates the Particularities of the New Architecture}

Today's information theory is based on the two principles documented  by Shannon and Wiener. In more detail, Shannon used the transition probability to describe the relation between the input, $\mathbf{x}$, and output $\mathbf{y}$, of an open-loop communication system, assuming that the channel response cannot be affected by external interventions. Similarly, Wiener examined a closed-loop communication system in which the receiver sends feedback to the transmitter in order for the later to optimize the nature of the input to the channel signal. In other words, both Shannon and Winner employed $\Pr\left(\mathbf{y}\left|\mathbf{x}\right.\right)$ as the fundamental pillar upon which the information theoretic framework was build. However, reconfigurable metasurface-coated devices have the capability of sensing the channel response and suitably manipulating the wireless environment. As a consequence, not only the nature of the system's input becomes customized, but also the environment itself. This contrasts the main assumption of both Shannon's and Wiener's principles and calls for a new information theoretic framework that is build upon $\Pr\left(\mathbf{y}\left|\mathbf{x}, \mathbf{F}\right.\right)$, where $\mathbf{F}$ stands for the reconfigurable metasurface response.  \\

\subsection{Transceivers and Transmission Waveforms Designs Optimization}

The use of metasurfaces for coating transceivers increases the transmission and reception directionality. To fully exploit the benefits of such designs, in directional wireless systems, the metasurface and transceivers' beamformers should be co-design in order to ensure that the transmitter beam aims at the direction in which the metasurface-coated receiver achieves the maximum gain. Moreover, to avoid severe performance degradation due to the operation of the receiver's low noise amplifiers beyond the saturation point, new power allocation schemes should be introduced that takes into account the existence or absence of such metasurfaces. Likewise,  metasurfaces have a spatial-dependent transfer function that limits the channel's coherent bandwidth. This characteristic aspires the mutual design of transmission waveforms and beamforming codebooks,  modulation schemes and low-weights channel~codes, as well as resource allocation and mobility management protocols. 

\section{Conclusions}
In this paper, we have introduced a new concept that completely differentiates from the classical reflection-based RIS-enabled communications. Instead, common passive metasurface-coats have been proposed, which have great potential to significantly increase the EE, EH and secure 6G communication systems. Electromagnetic-based simulation results have shown remarkable insights that can satisfy the future needs of communications systems both at the energy and security fronts.
\section*{Acknowledgment}
This work was supported by the National Natural Science Foundation of China under Grant No. 62071202 and 62171201.

\bibliographystyle{IEEEtran}
\bibliography{IEEEabrv,References}

\end{document}